\documentclass[rnote]{aa}
\usepackage{graphicx}
\usepackage{longtable}
\usepackage{txfonts}
\begin{document}
   \title{Chemical composition of A and F dwarfs members of the Pleiades open cluster\thanks{Based on observations performed at the Observatoire de Haute-Provence (France).}}


   \author{M. Gebran\inst{1} \and R. Monier\inst{2} }

   \offprints{M. Gebran}

   \institute{$^1$Groupe de Recherche en Astronomie et Astrophysique du Languedoc,UMR 5024, Universit\'e Montpellier II, Place Eug\`ene Bataillon, 34095 Montpellier, France. \email{gebran@graal.univ-montp2.fr} \\
   $^2$Laboratoire Universitaire d'Astrophysique de Nice, UMR 6525, Universit\'e de Nice - Sophia Antipolis, Parc Valrose, 06108 Nice Cedex 2, France. \email{Richard.Monier@unice.fr}}

   \date{Received; accepted}

 
  \abstract
   {}
   {Abundances of 18 chemical elements have been derived for 16 A (normal and chemically peculiar CP) and 5 F dwarfs members of the Pleiades open cluster in order to set constraints on evolutionary models.}
   {The abundances, rotational velocities and microturbulent velocities were derived by iteratively adjusting synthetic spectra to observations at high resolution (R $\simeq$ 42000 and R $\simeq$ 75000) and high signal-to-noise (S/N) ratios.}
   {The abundances obtained do not exhibit any clear correlation with the effective temperature nor the projected rotational velocity. Interestingly, A stars exhibit larger star-to-star variations in C, Sc, Ti, V, Cr, Mn, Sr, Y, Zr and Ba than F stars. F stars exhibit solar abundances for almost all the elements. In A stars, the abundances of Si, Ti and Cr are found to be correlated with that of Fe, the [X/Fe] ratios being solar for these three elements. \\
   The derived abundances have been compared to the predictions of published evolutionary models at the age of Pleiades (100 Myr). For the F stars, the predicted slight underabundances of light elements and overabundances of Cr, Fe and Ni are indeed confirmed by our findings. For A stars, the predicted overabundances in iron peak elements are confirmed in a few stars only. }
   {The large scatter of the abundances in A stars, already found in the Hyades, Coma Berenices and the UMa group and in field stars appears to be a characteristic property of dwarf A stars. The occurence of hydrodynamical processes competing with radiative diffusion in the radiative zones of the A dwarfs might account for the found scatter in abundances.}


\maketitle

\section{Introduction}

This paper is the second in a series about the abundances of 18 chemical elements in A and F dwarfs in open clusters of different ages. In a previous study, abundances for the same elements have been derived for 11 A and 11 F dwarfs members of the Coma Berenices open cluster by Gebran et al. (2007) (hereafter refered to as Paper I). The aims of this project are twofold: improve our knowledge of the chemical composition of A dwarfs (normal and CP stars) and set constraints on self-consistent evolutionary models including hydrodynamical and particle transport processes. At the age of Coma Berenices (about 450 Myr, \cite{1993A&A...268..829B}), the A stars have spent more time on the Main Sequence than at the age the Pleiades (around 100 Myr, \cite{1993A&AS...98..477M}). Comparison of the abundances derived for A and F stars in these 2 clusters should therefore help address the expected evolution of abundances with time predicted in the frame of the diffusion theory (Richer et al. 2000). \\
High-resolution spectroscopy of A and F stars in the Pleiades is feasible with 2-m class telescopes down to magnitude V=9 wich corresponds to the earliest F stars, the distance to this cluster being about 134$\pm$3 pc (\cite{2005A&A...429..887P}). In Table \ref{pleiades-biblio}, we have collected previous abundance determinations of A, F and G dwarfs in the Pleiades. Lithium abundances have been determined first by \cite{1987PASP...99.1288P} for A, F and G dwarfs then by \cite{2002A&A...391..253F} for G and K stars. Carbon and iron abundances were obtained for 12 F dwarfs by \cite{1990ApJ...351..480F} and \cite{1990ApJ...351..467B} respectively. They found a mean iron abundance of $\langle [\rm{Fe/H}] \rangle$=$-0.034\pm0.024$ dex. Beryllium abundances was derived by \cite{2003ApJ...582..410B} for 14 F and G dwarfs. \cite{1997A&A...318..870B} determined the abundances of Li, Al, Si, S, Fe, Ni and Eu for 5 normal A and 3 Am stars. Quite unexpectedly, they found the same iron abundance in A and Am stars, about twice the value determined from F stars. \cite{1998A&A...332..224H} determined the abundances of Mg, Ca, Sc, Cr, Fe and Ni for 9 A-Am stars of the Pleiades.\\
The main thrust of this paper is to report on the abundances of 18 chemical elements (C, O, Na, Mg, Si, Ca, Sc, Ti, V, Cr, Mn, Fe, Co, Ni, Sr, Y, Zr and Ba.) for 16 A and 5 F dwarfs in the Pleiades open cluster. As in Paper I, we have undertaken a search for correlations of the individual abundances with effective temperature (T$_{\rm{eff}}$), projected rotational velocity ($v_{e}\sin i$) and iron abundance. Any such correlation would be very valuable in theoretical investigations of photospheric abundances. The derived abundances were also compared to the predictions of self-consistent evolutionary models.

\begin{table*}[]
\centering
\label{pleiades-biblio}
\caption{Previous abundance determinations for the Pleiades A, F and G dwarfs.}
\begin{center}
\begin{tabular}{c c c c}
\hline
 & & \\
Reference  & Stars studied & Chemical Elements \\
\hline
 & & \\
 \cite{1987PASP...99.1288P}  & 18 A,F,G  & Li  \\
 \cite{1997A&A...318..870B}&  8 A-Am  & Li,Al,Si,S,Fe,Ni,Eu        \\
 \cite{1998A&A...332..224H} & 9 A-Am & Mg,Ca,Sc,Cr,Fe,Ni \\
  \hline \\
\cite{1990ApJ...351..467B} & 12 F    & Fe \\
\cite{1990ApJ...351..480F}       & 12  F   &   C  \\
\cite{2003ApJ...582..410B}           & 14 F and G  & Be \\
\cite{2002A&A...391..253F}& 11 G and K & Li \\	
\hline

\end{tabular}
\end{center}
\end{table*}

\begin{table*}[]
\caption{Data on the programme stars. Spectral type are taken from SIMBAD and WEBDA online database. $T_{\rm{eff}}$ and $\log g$ are those determined by UVBYBETA code. $v_{e}\sin i$  and $\xi_{t}$ are determined as explained in sect.\ref{abundances-analysis}. References (a), (b), (c) and (d) are \cite{1994A&AS..106..377C}, \cite{1972ApJ...176..367B}, \cite{1978PASP...90..201A} and \cite{1990BICDS..38..151R}.}
\label{etoiles-pleiades-vitesses}
\centering
\begin{tabular}{lcccccccc} 
\hline
HII &	 HD	& Type  & $m_{v}$       & $T_{\rm{eff}}$ 	    &  $\log g$     &	 $v_{e}\sin i$  & $\xi_{t}$ & Remarks \\
     &	   &	 	&		 &(K)		  		&	      &	km\,s$^{-1}$	         &km\,s$^{-1}$         &\\ \hline
  157     &   23157   &A5V   &7.95	   &7514		&4.26		&56		    &2.60	&  SB (a)     \\
  158     &   23156   &A7V   &8.23	   &7940		&4.23		&32.5		    &2.70	&  $\delta$ Scuti (b)	\\
  697     &   23375   &A9V   &8.58	   &7395		&4.22		&88		    &2.30	&	      \\
 1362     &   23607   &A7V   &8.25	   &8055		&4.32		&18.9		    &3.10	&  $\delta$ Scuti (b), Am (d)	      \\
 1397     &   23631   &A2V   &7.30	   &9613		&4.34		&7.5		    &2.10	&  SB, Am (c)	      \\
 1876     &   23763   &A1V   &6.96	   &8999		&4.19		&100		    &2.00	&	    \\
 2415     &   23924   &A7V   &8.10	   &8144		&4.29		&33.5		    &2.70	&	     \\    
 2488     &   23948   &A0    &7.54	   &9083		&4.35		&118		    &2.30	&	    \\
 5006     &   22615   &Am    &6.50	   &8407		&3.83		&29.5		    &4.00	&  $\notin$ Pleiades (d) \\
  531     &   23325   &Am    &8.57	   &7638		&4.23		&80		    &2.50	&	      \\
 1375     &   23629   &A0V   &6.28	   &9940	   &4.32	   &162 	   &1.55	   &	SB	   \\
 1380     &   23632   &A1V   &7.02	   &9616	   &4.22	   &200 	   &1.50	   &		   \\
 2195     &   23863   &A7V   &8.15	   &7911	   &4.10	   &157 	   &2.50	   &		   \\
 1028     &   23489   &A2V   &7.38	   &9078	   &4.25	   &120 	   &1.90	   &		   \\
 1993     &   23791   &A8V   &8.38	   &7796	   &4.32	   &75  	   &3.20	   &		   \\
  717     &   23387   &A1V   &7.19	   &9581	   &4.22	   &21  	   &0.50	   &	   SB	   \\ \hline \\ \hline
  605     &   23351   &F3V   &9.03	   &6863	   &4.38	   &14.8	   &1.45	   &		   \\
 1357     &   23609   &F8IV  &6.99	   &6492	   &4.28	   &9.8 	   &1.75	   &		   \\
  338     &   23247   &F3V   &9.06	   &6948	   &4.47	   &43.5	   &2.00	   &		   \\
 1766     &   23732   &F4V   &9.21	   &6837	   &4.54	   &25  	   &1.70	   &		   \\
 1122     &   23511   &F4V   &9.28	   &6730	   &4.63	   &30.5	   &1.60	   &		   \\

\hline	
\end{tabular}
 \end{table*}

\section{Program stars, observations and data reduction}
 Twenty one stars members of the Pleiades cluster were observed in January 2004 and in December 2006. We selected stars evenly sampling in mass the Main Sequence. The selected 15 A stars amount to about half the total number of A dwarfs in the Pleiades. These stars were observed using ELODIE and SOPHIE, two \'echelle spectrographs at the Observatoire de Haute-Provence (OHP). ELODIE is a fiber-fed cross-dispersed echelle spectrograph attached to the 1.93-m telescope at OHP (\cite{1996A&AS..119..373B}). An ELODIE spectrum extends from 3850 to 6811 \AA \ at a resolving power of about 42000. ELODIE was replaced by SOPHIE in September 2006. SOPHIE spectra cover the wavelength interval from 3820 to 6930 \AA \ in 39 orders with two different spectral resolutions: the high resolution mode HR (R=75000) and the hight efficiency mode HE (R=39000). SOPHIE being about 2 magnitudes more sensitive in the V band than ELODIE, we could obtain spectra in the HR mode with signal to noise ratios ranging from 100 to 300 in less than 75 minutes. The five early F stars being fainter than V=9 mag were observed with SOPHIE. \\
The basic data of these stars are collected in Table \ref{etoiles-pleiades-vitesses}. The Hertzsprung and Henry Draper identifications appear in columns 1 and 2. In column 3 we gather the spectral type and in column 4 the apparent visual magnitude. Effective temperatures (T$_{\rm{eff}}$) and surface gravities ($\log g$) deduced from the $uvby\beta$ photometry are collected in columns 5 and 6. The derived projected rotational velocities and microturbulence velocities are collected in columns 7 and 8. Comments about the membership, binarity and pulsation appear in the last column. The rotational velocities of these stars were found to range from 7.5 km\,s$^{-1}$ to 200 km\,s$^{-1}$, six of the A stars are rotating with a $v_{e}\sin i$ larger than 100 km\,s$^{-1}$. According to the CCDM catalogue (Dommanget \& Nys 1995), HD23387 is the primary in a spectroscopic binary. Its companion located at 0.3 arcsec, has a visual magnitude V=9 and should contribute about 20\% of the light in the spectrum. The spectral type of this companion is unknown. Careful inspection of the spectrum of HD23387 does not reveal lines which could be attributed to a companion of a later spectral type than that of the primary. The abundances deduced for HD23387 should be taken with caution.\\
The reduction of ELODIE's spectra was fully explained in Paper I and follows the method of \cite{2002A&A...383..227E}. A similar reduction procedure was applied to SOPHIE spectra. The correction for scattered light was found to be about 1\% in the blue part of the spectrum and less elsewhere.

\section{Abundance analysis}
\label{abundances-analysis}
\subsection{Method and input data}
The most appropriate method to derive abundances is spectrum synthesis as several of the investigated stars are fast rotators. The iterative method of Takeda (1995) was used to derive abundances of 18 chemical elements. The method consists in adjusting iteratively synthetic spectra to the observed normalized spectra and minimizing the chisquare between the models and the observations (see Gebran et al. 2007 for a detailed discussion).\\
The effective temperatures and surface gravities were determined using \cite{1993A&A...268..653N} UVBYBETA code. This code is based on \cite{1985MNRAS.217..305M} grid which calibrates the $uvby\beta$ photometry in terms of T$_{\rm{eff}}$ and $\log g$. The photometric data were taken from \cite{1998A&AS..129..431H}. The estimated errors on T$_{\rm{eff}}$ and $\log g$ are $\pm$125 K and $\pm$0.20 dex respectively. The derived effective temperatures and surface gravities are collected in columns 5 and 6 of Table \ref{etoiles-pleiades-vitesses}.\\
LTE model atmospheres were calculated using Kurucz's ATLAS9 code (\cite{1992RMxAA..23..181K}). The solar abundances used in ATLAS9 and in the synthetic spectra computation are those from \cite{1998SSRv...85..161G}. This version of ATLAS9 uses the new opacity distribution function (ODF) (\cite{{2003IAUS..210P.A20C}}). The microturbulent velocity is constant with depth and was adopted as prescribed by Smalley (2004). For stars with $T_{\rm{eff}}$ $\leq$ 8500 K, convection was taken into account in the model computations with a mixing-length ratio equal to 1.25. \\
We used the same linelist as in Paper I (270 transitions for 18 elements). The LTE assumption is justified for most of the lines studied because they are weak and are formed deep in the atmospheres where LTE should hold. A few lines of CI, MgII and BaII analysed here are affected by non-LTE effects (Gebran et al. 2007). The derived abundances for these elements appear in brackets in the online tables \ref{abondances-A} and \ref{abondances-F}.

\subsection{Results}
Projected rotational velocities, microturbulent velocities and abundances of 18 chemical elements were determined following the same procedure as in Paper I (section 3.2). The rotational and microturbulent velocities were determined by using a set of weak and strong unblended lines of iron and the MgII triplet at $\lambda$4481 \AA \ (section 3.2.1 of Paper I). The derived rotational velocities agree well with those in \cite{1998A&A...332..224H} for seven stars in common to both studies. The microturbulent velocities were also found to comply with the prescription of \cite{2004IAUS..224..131S}. In order to validate the abundances, we double-cheked them using the code SYNSPEC48 (\cite{1992A&A...262..501H}). The abundances derived using Takeda's procedure were found to agree within their uncertainties with those derived from SYNSPEC48. The abundances presented in the online tables (\ref{abondances-A} and \ref{abondances-F}) are those derived from Takeda's iterative procedure. The errors on abundances were computed as explained in the appendix of Paper I. 

\section{Discussion and conclusion}
\subsection{Search for correlations with stellar parameters}
\begin{figure*}[]
\centering
\includegraphics[scale=0.34]{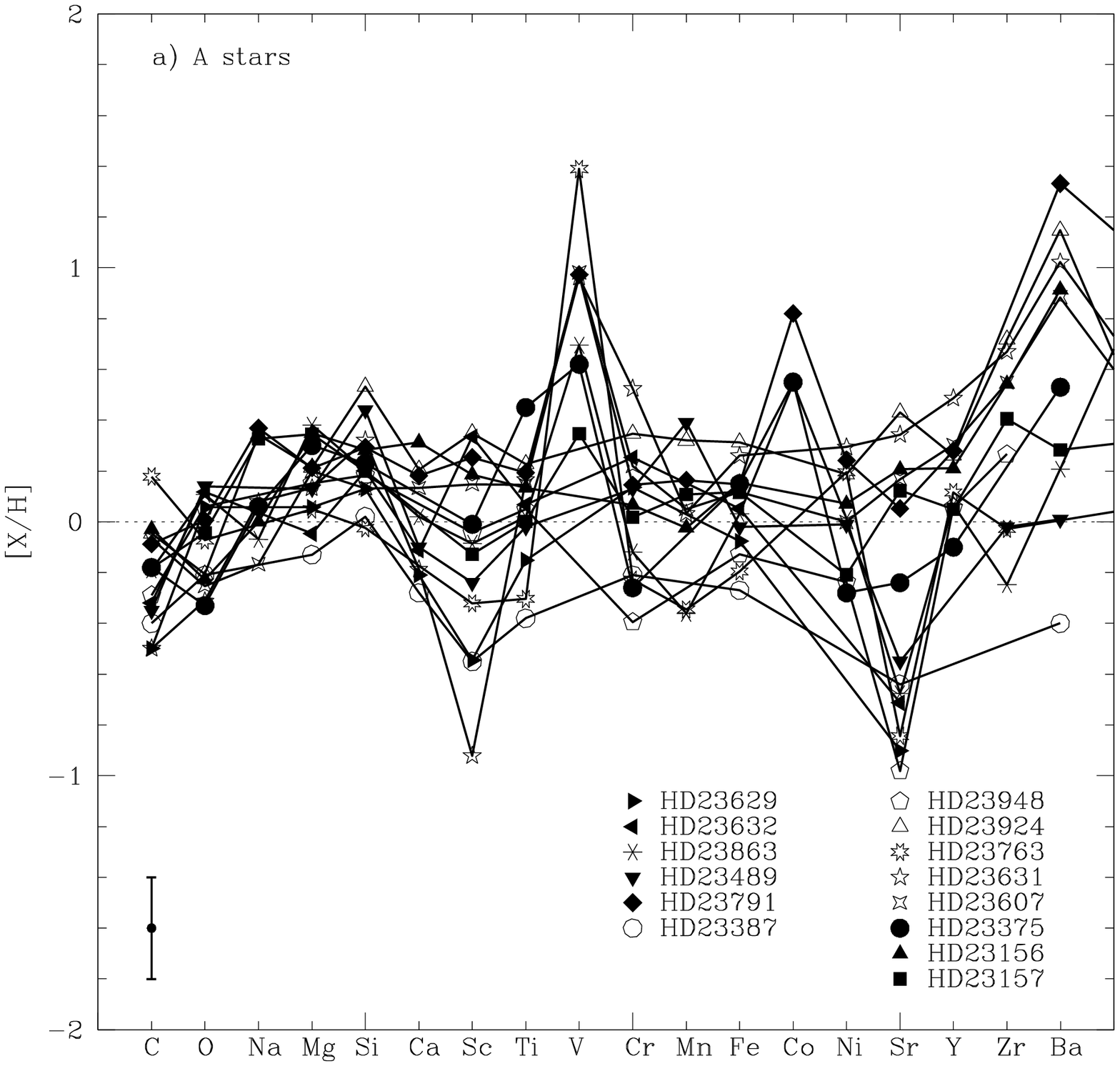} \ \ 
\includegraphics[scale=0.34]{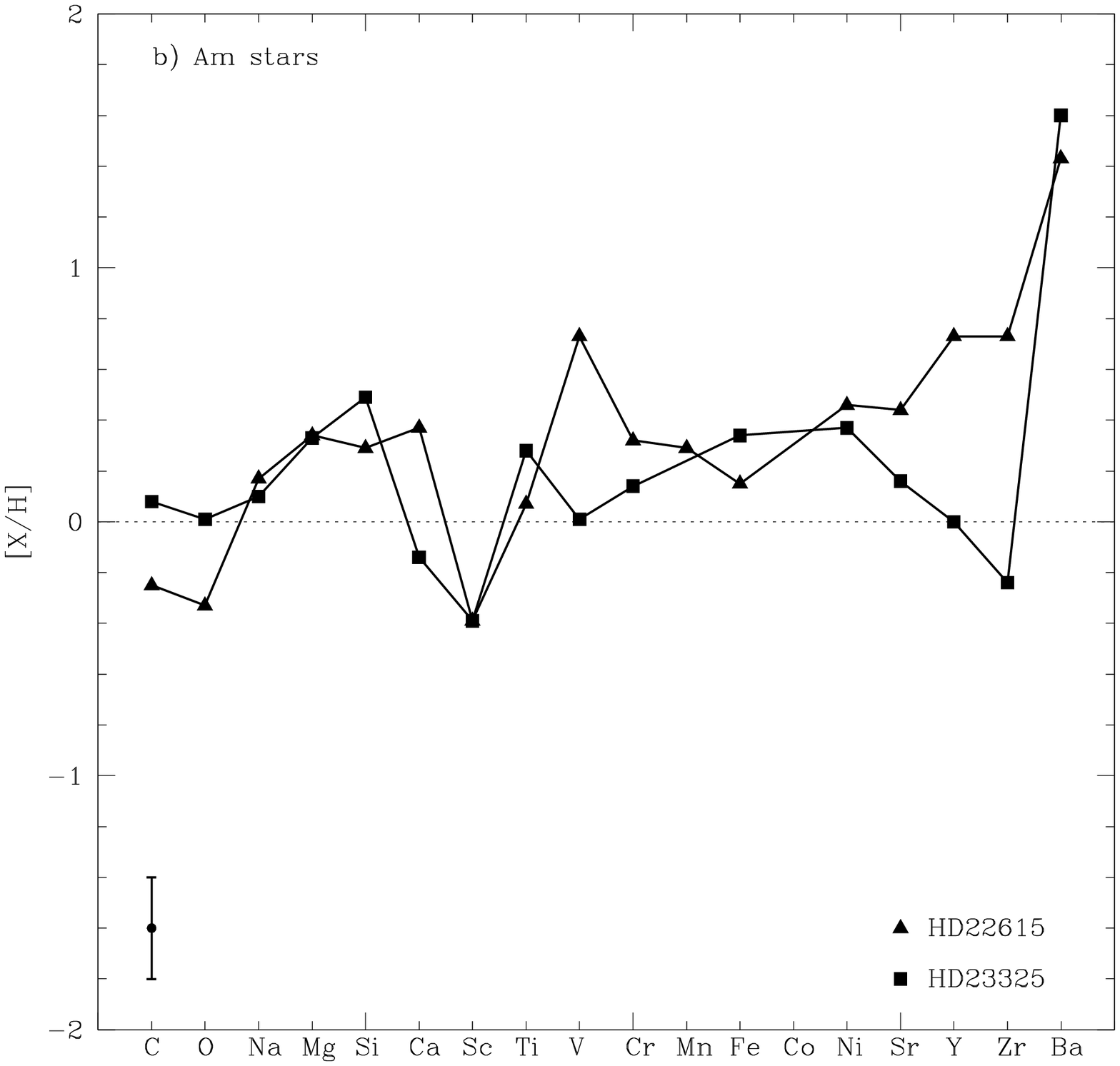}
\caption{Abundance patterns of normal A (a) and Am (b) stars members of Pleiades Cluster. The horizontal dotted line represents the solar value.}
\label{Astars}
 \end{figure*}

\begin{figure}[]
\centering
\includegraphics[scale=0.35]{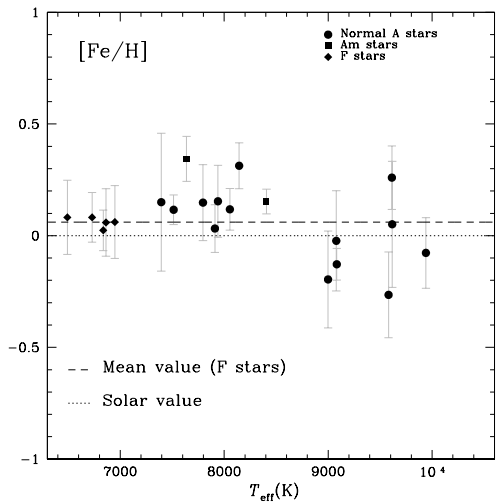}  
\caption{[Fe/H] versus $T_{\rm{eff}}$ for the 21 stars. The dotted line represents the solar value and the dashed line represents the mean abundance of iron for the F dwarfs. F stars are depicted as diamonds, Am stars as squares and normal A stars as circles.}
\label{Fe-Teff}
 \end{figure}

The behaviour of the abundances with respect to stellar parameters (effective temperature, apparent rotational velocity and also iron abundance) was investigated. No convincing correlation of any of the derived abundances was found either with $T_{\rm{eff}}$ nor $v_{e}\sin i$. However, the most distinctive finding is large star-to-star variations for several of the abundances among the A stars, in particular for C, Sc, Ti, Cr, Fe, Sr, Y, Zr and Ba. This behaviour is conspicuous in Figure \ref{Astars} where the abundances of all chemical elements are displayed for all A stars. This is also readily seen in graphs where [X/H] is displayed versus effective temperature: in such graphs, A stars display abundances which are more scattered around their mean value than the F stars do. The case of iron is examplified in Figure \ref{Fe-Teff}. Similar star-to-star variations in the abundances of several chemical elements were already found in A stars in other clusters or moving groups: in the Hyades (\cite{varenne-monier-99}), in the UMa group (\cite{monier2005}) and in the Coma cluster (Gebran et al. 2007).\\
The two Am stars (HD23325 and HD22615) analysed in this study are both underabundant in scandium. Calcium is underabundant in HD23325 while it is overabundant in HD22615 by about $\sim$0.37 dex. Cr, Fe, Ni, Sr and Ba are enhanced in these Am stars. We confirm that HD23631 (A2V) should be classified as an Am star as already suggested by Conti (1968), Gray \& Garrison (1987) and Renson (1990). Indeed, we found for this star an apparent rotational velocity of 7.5 km\,s$^{-1}$, deficiencies in calcium and scandium ($-$0.19 dex and $-$0.92 dex respectively) and enrichment in iron peak elements and heavy elements. HD23631 was found to be a short period (P=7.34 days) binary Am stars by \cite{1968AJ.....73..348C}. \cite{1987ApJS...65..581G} have also classified HD23631 as an A0mA1Va using intermediate resolution spectra. \\
In normal A stars, the ratios [C/Fe] and [O/Fe] are found to be anticorrelated with [Fe/H] as is the case in Coma Berenices. The abundances of Mg, Si, Ti and Cr are found to be correlated with that of Fe (correlation coefficients are 0.63, 0.78, 0.84 and 0.65 respectively). The ratios [X/Fe] for these elements are close to solar as already found for field A stars for Si and Ti (Lemke 1989, Hill and Landstreet 1993).\\
Concerning F stars, we found a mean iron abundance of $\langle [\rm{Fe/H}] \rangle$=0.06$\pm$0.02 dex based on the 5 stars of our sample using FeII lines. This value is slightly larger than that derived by Boesgaard \& Friel (1990) in their analysis of 12 F stars: -0.034 dex based on their analysis of 15 Fe I lines. The difference might be accounted for by the usage of different lines and also of different effective temperatures and surface gravities.

\subsection{Evolutionary models} 
The derived abundances can also be compared to the predictions of evolutionary models. These abundances are very useful to constrain the physics included in the code. For F stars, the models calculated by \cite{1998ApJ...504..559T} include radiative diffusion and gravitational settling for 28 chemical elements (Z $\leq$ 28). At the age of the Pleiades cluster (100 Myr), these models predict slight underabundances of carbon and oxygen for stars with $T_{\rm{eff}} > 6500$ K which are confirmed in our data. However the predicted underabundances of Mg, Si and Ca are not found in our analysis. The predicted slight overabundances for Cr, Fe and Ni are indeed observed for the five F stars analysed. The overall trend for F stars, slight underabundances of light elements and overabundances of iron-peak elements agree well with the predictions of the models (except for a few elements) suggesting that the appropriate physics is included in the code for F stars.\\
For A stars, we can compare our results to the predictions of the Richer et al. (2000) models which include turbulent diffusion in addition to radiative diffusion. Inspection of their figures 10 and 11 reveals that at the age of the Pleiades, C and O should be slightly underabundant by about $-$0.1 dex and the iron peak elements be slightly overabundant from about 0.1 to 0.4 dex. The derived abundances respectively for C, O and iron peak elements in this analysis agree well with these levels of deficiency and enrichment.\\
The star-to-star scatter of abundances among the 16 A stars analysed here was already previously found in other open clusters and in a few field A stars. It appears to be a characteristic property of dwarf A and F stars and strongly suggests that hydrodynamical processes competing with radiative and turbulent diffusion must be at work in the radiative zones of these stars (see a review in Zahn 2005). In order to fully characterize this scatter, spectroscopy of the remaining A and F stars in the Pleiades is foreseen in the near future. Also non-LTE analysis should be performed to yield more accurate abundances for C, Mg and Ba as most of the lines analysed here for these elements are prone to non-LTE effects.

\Online
 
 \begin{table*}
\small
\caption{Abundances relative to hydrogen and to the solar value, $[\frac{X}{H}]=\log(\frac{X}{H})_{\star}-\log(\frac{X}{H})_{\odot}$ for the A stars. The solar values are those of \cite{1998SSRv...85..161G}. The HD numbers in italic are those for which the uncertainties have been calculated as explained in Appendix A of Paper I. For the others, the quantities labeled as $\sigma$ are standard deviations. The lines of the elements in brackets should be treated in the non-LTE case.  }
\label{abondances-A}
\centering 
 \begin{tabular}{cccccccccccc} 
\hline
HD   & SpT 	& $<$CI$> $	 &$\sigma_{C}$	 & OI   &$\sigma_{O}$& NaI   &$\sigma_{Na}$	& $<$MgII$> $  &$\sigma_{Mg}$	&SiII&$\sigma_{Si}$ \\ \hline
\it{HD23157}&A5V  &-0.18  & 0.05	& -0.04  & 0.14  &  0.33   & 0.06   & 0.34   & 0.32   & 0.20	&0.09	 \\
HD23156&A7V  &-0.03  & 0.07	& -0.23  & 0.20  &  0.00   & 0.11   & 0.35   & 0.32   & 0.28	&0.19	 \\
\it{HD23325}&Am   &0.08   & 0.04	& 0.01   & 0.09  &  0.10   & 0.16   & 0.33   & 0.16   & 0.49	&0.11	 \\
HD23375&A9V  &-0.18  & 0.40	& -0.33  & 0.14  &  0.06   & 0.27   & 0.30   & 0.32   & 0.23	&0.33	 \\
HD23607&A7V  &-0.07  & 0.10	& -0.25  & 0.01  &  -0.17  & 0.21   & 0.20   & 0.32   & 0.13	&0.07	 \\
HD23631&A2V  &-0.50  & 0.04	& -0.31  & 0.05  &  0.08   & 0.12   & 0.13   & 0.18   & 0.32	&0.10	 \\
HD23763&A1V  &0.18   & 0.31	& -0.07  & 0.06  &-	   &-	    & 0.05   & 0.21   & -0.02	&0.16	 \\
HD23924&A7V  &-0.05  & 0.11	& -0.21  & 0.23  &  0.34   & 0.21   & 0.21   & 0.30   & 0.53	&0.09	 \\
\it{HD23948}&A0   &-0.29  & 0.07	& 0.06   & 0.06  &-	   &-	    & 0.15   & 0.09   & 0.18	&0.13	 \\
\it{HD22615}&Am   &-0.25  & 0.08	& -0.22  & 0.06  &  0.17   & 0.06   & 0.34   & 0.13   & 0.29	&0.08	 \\
HD23629&A0V  &-0.50  & 0.10	& 0.06   & 0.06  &-	   &-	    & 0.06   & 0.34   & 0.13	&0.08	 \\
HD23632&A1V  &-0.32  & 0.11	& 0.12   & 0.01  &-	   &-	    & -0.05  & 0.20   & 0.20	&0.22	 \\
HD23863&A7V  &-0.19  & 0.11	& 0.11   & 0.05  &  -0.07  & 0.19   & 0.38   & 0.32   &-  	&-	 \\
HD23489&A2V  &-0.35  & 0.08	& 0.14   & 0.06  &-	   &-	    & 0.13   & 0.15   & 0.44	&0.26	 \\
HD23791&A8V  &-0.09  & 0.26	& 0.00   & 0.14  &  0.37   & 0.08   & 0.21   & 0.12   & 0.29	&0.04	 \\
HD23387&A1V  &-0.40  & 0.05	& -0.21  & 0.04  &-	   &-	    & -0.13  & 0.09   & 0.02	&0.09	 \\

\hline \hline

HD   & SpT &CaII   &$\sigma_{Ca}$ &ScII	 &$\sigma_{Sc}$&TiII&$\sigma_{Ti}$&VII   &$\sigma_{V}$&CrII 	&$\sigma_{Cr}$	\\ \hline
\it{HD23157}&A5V  &-   	 &-	  &  -0.13   & 0.11  &  0.00	 &0.11   &  0.35  &   0.46  &	0.02	 &0.09 	\\	    
HD23156&A7V  &   0.31	 & 0.12   &  0.19    & 0.22  &  0.13	 &0.13   &-	  &-	    &	0.07	 &0.09	\\
\it{HD23325}&Am   &   -0.14   & 0.30   &  -0.39   & 0.22  &  0.28	 &0.15   &  0.01  &   0.29  &	0.14	 &0.20	\\
HD23375&A9V  & -  	 &-	  &  -0.01   & 0.14  &  0.45	 &0.24   &-	  &-	    &	-0.26	 &0.28	\\
HD23607&A7V  &   0.13	 & 0.44   &  0.15    & 0.10  &  0.14	 &0.25   &  0.98  &   0.03  &	0.22	 &0.21	\\
HD23631&A2V  &   -0.19   & 0.05   &  -0.92   & 0.05  &  0.16	 &0.22   &  0.96  &   0.06  &	0.52	 &0.17	\\
HD23763&A1V  &  - 	 &-	  &  -0.32   & 0.33  &  -0.30	 &0.15   &  1.39  &   0.08  &	-0.23	 &0.28	\\
HD23924&A7V  &   0.21	 & 0.22   &  0.35    & 0.17  &  0.23	 &0.22   &-	  &-	    &	0.35	 &0.08	\\
\it{HD23948}&A0   &   -	 &-	  &  -0.04   & 0.13  &  0.05	 &0.08   &-	  &-	    &	-0.39	 &0.21	\\
\it{HD22615}&Am   &   0.37	 & 0.10   &  -0.39   & 0.10  &  0.07	 &0.07   &  0.96  &   0.12  &	0.30	 &0.10	\\
HD23629&A0V  &   -0.21   & 0.09   &  -0.54   & 0.34  &  -0.15	 &0.13   &-	  &-	    &	0.14	 &0.17	\\
HD23632&A1V  &   -0.11   & 0.04   &  0.34    & 0.09  &  0.07	 &0.22   &-	  &-	    &	0.25	 &0.14	\\
HD23863&A7V  &	 0.02	 & 0.40   &  -0.08   & 0.21  &  0.02	 &0.20   &  0.70  &   0.11  &	-0.12	 &0.36	\\
HD23489&A2V  &	 -0.10   & 0.14   &  -0.24   & 0.26  &  -0.02	 &0.18   &-	  &-	    &	0.13	 &0.18	\\
HD23791&A8V  &	 0.18	 & 0.06   &  0.25    & 0.42  &  0.19	 &0.22   &  0.97  &   0.18  &	0.15	 &0.21	\\
HD23387&A1V  &	 -0.28   & 0.14   &  -0.55   & 0.09  &  -0.38	 &0.17   &-	  &-	    &	-0.21	 &0.12	\\

\hline \hline
HD   & SpT &MnI   &$\sigma_{Mn}$ &FeII	 &$\sigma_{Fe}$&CoI   &$\sigma_{Co}$&NiI 	&$\sigma_{Ni}$& SrII&$\sigma_{Sr}$	\\ \hline
\it{HD23157}&A5V  &  0.11    &  0.20  &    0.12    &  0.07  &- 	   &- 	      &   -0.21   &  0.10   &   0.12   &   0.34  \\		      
HD23156&A7V  &  -0.02   &  0.08  &    0.15    &  0.16  &-   	   &-	      &   0.07    &  0.15   &   0.21   &   0.07  \\
\it{HD23325}&Am   &-         &  -	 &    0.34    &  0.10  &-   	   & -	      &   0.37    &  0.17   &   0.16   &   0.26  \\
HD23375&A9V  &-         &  -	 &    0.15    &  0.31  &    0.55   &   0.08   &   -0.28   &  0.33   &   -0.24  &   0.07  \\
HD23607&A7V  &  0.05    &  0.12  &    0.12    &  0.09  &-   	   &- 	      &   0.00    &  0.16   &   0.17   &   0.01  \\
HD23631&A2V  &  0.03    &  0.08  &    0.26    &  0.14  &-   	   &- 	      &   0.29    &  0.11   &   0.34   &   0.09  \\
HD23763&A1V  &  -0.35   &  0.37  &    -0.20   &  0.22  &-   	   &- 	      &   0.19    &  0.20   &   -0.84  &   0.34  \\
HD23924&A7V  &  0.32    &  0.09  &    0.31    &  0.10  &-   	   &-	      &   0.19    &  0.15   &   0.43   &   0.15  \\
\it{HD23948}&A0   &-         &  -	 &    -0.13   &  0.07  &-   	   &-	      &   -0.24   &  0.27   &   -0.98  &   0.24  \\
\it{HD22615}&Am   &  0.29    &  0.10  &    0.15    &  0.06  &-   	   &- 	      &   0.46    &  0.11   &   0.44   &   0.30  \\
HD23629&A0V  &-         & - 	 &    -0.08   &  0.16  &-   	   &-	      & -   	  &-	    &   -0.90  &   0.05  \\
HD23632&A1V  &-         &-  	 &    0.05    &  0.28  &-   	   &-	      &  -  	  &-	    &   -0.71  &   0.20  \\
HD23863&A7V  &	-0.36   &  0.27  &    0.03    &  0.11  &    0.54   &   0.08   &   0.01    &  0.22   &   -0.68  &   0.39  \\
HD23489&A2V  &	0.39    &  0.19  &    -0.02   &  0.22  &-   	   &-	      &   -0.01   &  0.24   &   -0.55  &   0.23  \\
HD23791&A8V  &	0.16    &  0.17  &    0.15    &  0.17  &    0.82   &   0.08   &   0.24    &  0.34   &   0.05   &   0.10  \\
HD23387&A1V  &-         &-  	 &    -0.27   &  0.19  & -  	   &-	      &  -  	  &-	    &   -0.64  &   0.04  \\
\hline \hline
HD   & SpT &YII   &$\sigma_{Y}$ &ZrII	 &$\sigma_{Zr}$&$<$BaII$>$   &$\sigma_{Ba}$& & &   & 	\\ \hline
\it{HD23157}&A5V  &  0.05  &   0.11 &    0.41  &   0.14   &  0.28&     0.16   &  &   &   &  \\		  
HD23156&A7V  &  0.21  &   0.05 &    0.54  &   0.05   &  0.91&     0.14   & 	  & 	    & 	  	& 	\\  
\it{HD23325}&Am   &  0.00  &   0.18 &    -0.24 &   0.24   &  1.60&     0.17   &  & &  	& 	\\  
HD23375&A9V  &  -0.10 &   0.13 &- 	  & -	     &  0.53&     0.04   &  & 	    & 	  	& 	\\  
HD23607&A7V  &  0.30  &   0.09 &    0.55  &   0.14   &  0.88&     0.13   &   &     &   	&  \\  
HD23631&A2V  &  0.49  &   0.06 &    0.67  &   0.11   &  1.02&     0.01   &    &    &   	& 	\\  
HD23763&A1V  &  0.12  &   0.36 &    -0.03 &   0.11   &      & -	  	 &  &     &  	&   \\  
HD23924&A7V  &  0.26  &   0.17 &    0.72  &   0.10   &  1.15&     0.27   &   &     &  	& \\  
\it{HD23948}&A0   &  0.06  &   0.14 &    0.27  &   0.25   &      & -	  	 &  &    & 	& 	\\  
\it{HD22615}&Am   &  0.73  &   0.13 &    0.73  &   0.16   &  1.43&     0.18   &   &     &   	&  \\  
HD23629&A0V  &-       &   -    &-  	  &- 	     &      & -	  	 && &	&  \\  
HD23632&A1V  &-       &  -     &-  	  &- 	     &      & -	  	 && &	&  \\
HD23863&A7V  &	0.07  &   0.14 &    -0.25 &   0.05   &  0.21&     0.28   && &	&  \\
HD23489&A2V  &-       & -      &    -0.02 &   0.16   &  0.01&     0.17   && &	&  \\  
HD23791&A8V  &	0.28  &   0.26 & -	  & -	     &  1.33&     0.14   && &	&  \\  
HD23387&A1V  &-	      &-       &  -	  &- 	     &  -0.40&    0.17   && &	&  \\  
\hline \hline

\end{tabular}
 \end{table*}

\begin{table*}
\caption{Abundances relative to hydrogen and to the solar value, $[\frac{X}{H}]=\log(\frac{X}{H})_{\star}-\log(\frac{X}{H})_{\odot}$ for the F stars. The HD numbers in italic are those for which the uncertainties have been calculated as explained in Appendix A of Paper I. For the others, the quantities labeled as $\sigma$ are standard deviations. The lines of the elements in brackets should be treated in the non-LTE case.}
\label{abondances-F}
\centering 
 \begin{tabular}{cccccccccccc} 
\hline
HD   & SpT 	& $<$CI$>$ 	 &$\sigma_{C}$	 & OI   &$\sigma_{O}$& NaI   &$\sigma_{Na}$	& $<$MgII$>$   &$\sigma_{Mg}$	&SiII&$\sigma_{Si}$ \\ \hline
HD23351&F3V  &-0.06  & 0.12	& -0.34  & 0.18  &  -0.20  & 0.16   & 0.14   & 0.36   & 0.05	&0.10	 \\
HD23609&F8IV &0.00   & 0.13	& -0.26  & 0.04  &  0.22   & 0.09   & 0.08   & 0.36   & 0.23	&0.07	 \\
HD23247&F3V  &-0.07  & 0.12	& -0.03  & 0.15  &  -0.14  & 0.17   & 0.43   & 0.36   & 0.35	&0.06	 \\
HD23732&F4V  &-0.06  & 0.06	& 0.08   & 0.32  &  -0.15  & 0.14   & 0.21   & 0.36   & 0.16	&0.12	 \\
HD23511&F4V  &-0.06  & 0.08	& -0.01  & 0.38  &  -0.03  & 0.30   & 0.33   & 0.36   & 0.34	&0.09	 \\

\hline \hline

HD   & SpT &CaII   &$\sigma_{Ca}$ &ScII	 &$\sigma_{Sc}$&TiII&$\sigma_{Ti}$&VII   &$\sigma_{V}$&CrII 	&$\sigma_{Cr}$	\\ \hline
HD23351&F3V  &   0.00	 & 0.10   &  -0.01   & 0.22  &  0.07	 &0.21   &  0.30  &   0.13  &	0.03	 &0.13	\\
HD23609&F8IV &   0.36	 & 0.20   &  0.14    & 0.11  &  0.06	 &0.14   &  0.64  &   0.26  &	0.15	 &0.09	\\
HD23247&F3V  &   0.00	 & 0.02   &  -0.05   & 0.22  &  0.12	 &0.25   &  0.57  &   0.27  &	0.04	 &0.12	\\
HD23732&F4V  &   0.31	 & 0.13   &  -0.05   & 0.21  &  0.06	 &0.21   &  0.44  &   0.39  &	0.02	 &0.15	\\
HD23511&F4V  &	 0.33	 & 0.17   &  0.06    & 0.27  &  0.14	 &0.22   &  0.35  &   0.32  &	0.08	 &0.14	\\

\hline \hline
HD   & SpT &MnI   &$\sigma_{Mn}$ &FeII	 &$\sigma_{Fe}$&CoI   &$\sigma_{Co}$&NiI 	&$\sigma_{Ni}$& SrII&$\sigma_{Sr}$	\\ \hline
HD23351&F3V  &  0.02    &  0.13  &    0.06    &  0.15  &    -0.21  &   0.23   &   -0.01   &  0.24   &   0.03   &   0.00  \\
HD23609&F8IV &  0.31    &  0.21  &    0.08    &  0.17  &    0.33   &   0.15   &   0.27    &  0.12   &   0.06   &   0.05  \\
HD23247&F3V  &  -0.12   &  0.14  &    0.06    &  0.16  &    -0.01  &   0.57   &   -0.04   &  0.18   &   0.09   &   0.05  \\
HD23732&F4V  &  0.13    &  0.16  &    0.02    &  0.09  &    0.17   &   0.14   &   0.03    &  0.20   &   0.06   &   0.03  \\
HD23511&F4V  &	0.23    &  0.20  &    0.08    &  0.11  &    -0.30  &   0.24   &   0.20    &  0.22   &   0.12   &   0.04  \\
\hline \hline
HD   & SpT &YII   &$\sigma_{Y}$ &ZrII	 &$\sigma_{Zr}$&$<$BaII$>$   &$\sigma_{Ba}$&   &  & & 	\\ \hline
HD23351&F3V  &  0.11  &   0.08 &    0.19  &   0.20   &  0.83&     0.22   & &  &  &	  \\
HD23609&F8IV &  0.14  &   0.08 &    0.22  &   0.13   &  0.74&     0.22   & &  &  &	  \\
HD23247&F3V  &  -0.02 &   0.11 &    0.37  &   0.20   &  0.66&     0.21   & &  &  &	  \\
HD23732&F4V  &  0.06  &   0.12 &    0.31  &   0.13   &  0.76&     0.23   & &  &  &	  \\
HD23511&F4V  &	0.11  &   0.16 &    0.65  &   0.34   &  0.80&     0.25   & &  &  &	  \\
\hline \hline

\end{tabular}
 \end{table*}

\end{document}